# What UAE Software Students Think about Software Testing: A Replicated Study

Luiz Fernando Capretz[1] [0000-0001-6966-2369] and Saad Harous[2] and Ali Bou Nassif [2]

[1] Western University, London, N6G5H5, Canada
`lcapretz@uwo.ca`
[2] University of Sharjah, Sharjah, United Arab Emirates
`harous@sharjah.ac.ae`
[2] University of Sharjah, Sharjah, United Arab Emirates
`anassif@sharjah.ac.ae`

**Abstract.** Software testing is vital to improve software quality. However, software tester role is stigmatized, partly due to misperception and partly due to the treatment of the testing process within the software industry. The present study analyses this situation aiming to explore what might inhibit an individual from taking up a software testing career. In order to investigate this issue, we surveyed 132 senior students pursuing degrees in information systems, information and communication technology, computer science, computer engineering, software engineering, and other closely-related disciplines at three universities in the United Arab Emirates: two publicly funded and one top-notch private university. The students were asked to describe the PROs and CONs of taking up a career in software testing and to ponder the likelihood that they would take up the career themselves. The study identified 7 main PROs and 9 main CONs for pursuing a testing career, and indicated that the role of software tester is perceived as a social role, which may require as many soft skills as technical prowess. The results also show that UAE software-related students have a stronger negative attitude towards software testing compared to their counterparts in other countries where similar investigations have been carried out in the past three years.

## 1   Introduction

As software systems are becoming more pervasive, they are also becoming susceptible to failures, resulting in potentially lethal combinations. There have been catastrophic failures such as Therac-25 [1], Ariane 5 [2], and most recently the Boeing 737-Max crisis [3], which led to the loss of capital and life. Many more similar incidents are happening now all over the world as we write this article.

Software testing is critical to prevent software failures. Therefore, research has indeed been carried out in testing, but these works are largely limited to the process [4] and technology [5], [6] dimensions; the human dimension has not been sufficiently addressed. Even though there are reports about inadequacies of testing professionals and their skills [7], only a few studies have tackled the education and curriculum dimensions directly [9].

The role of tester does not appear as one of the preferred roles among the population of software practitioners, according to previous study results [9] [10] [11]. Some studies point out the need for reversing people´s perceptions regarding this role by using career progression [12] and other incentives to reinforce the crucial insights that a tester brings to the software project [13].

It has been pointed out that human and social aspects play a significant role in software testing practices [14] [15]. Attention to human factors in software testing in an academic setting has been encouraged by Hazzan and Tomayko [16], Shah and Harold [17], and Capretz [18]. Therefore, we decided to explore the human dimension of this endeavor. We started with the basic problem that plagues the testing profession, the shortage of talent, by asking why students are reluctant to consider testing careers.

If very few bright individuals voluntarily choose testing careers, this robs the industry of good testing and delivery of quality products. To change this situation, it is necessary to analyze the reasons for such apathy towards testing careers.

In this research, we studied the chances of senior software-related degree students taking up software testing careers and their reasons. To that end, we conducted surveys with senior software-related programs at three universities in the United Arab Emirates endeavoring to expose actual reactions of students to the role of tester in the software industry.

## 2    Methodology

The objectives of the study were explained to students, and their responses to the survey were sought. They were assured that their responses would not influence course grades in any way and were offered an option of not disclosing gender and GPA. A fresh and simple survey questionnaire, which is presented in Appendix A. The first author conducted replicated research using the same questionnaire in Canada, India, China, Cuba, Pakistan, Malaysia and Brazil, thus ascertained that the students neither had any queries nor reported any lack of clarity in the questions.

The first two questions were open ended questions: 1) What are three PROs (in order of importance) of pursuing a career in software testing; and 2) What are three CONs (also prioritized) of pursuing a career in software testing. The third question asked participants to indicate their intentions of pursuing a career in software testing. They were given the option to answer with either "certainly not," "no," "maybe," "yes," and "certainly yes." Participants were also invited to share the reasons behind their responses. Lastly, participants were asked to provide demographic information about themselves.

The research is descriptive, diagnostic, cross-sectional, and mixed. Descriptive research describes the characteristics of a population being studied and does not explore the reasons for those characteristics. Diagnostic research studies determine the frequency with which something occurs or its association with something else. We studied the event not over time but at a cross-section, making the study cross-sectional. We used a qualitative method of seeking open-ended responses about the PROs and CONs of testing careers and quantitative method, we asked categorical questions about the probability of choosing testing careers, making the study a mixed one too.

Our study analyzed the reasons that UAE students did not choose testing careers. These students are seniors pursuing a degree in information systems, information and communication technology, computer science, computer engineering, software engineering, and other closely-related disciplines at three universities in the UAE. One hundred and thirty two (132) students voluntarily participated in the survey during the Fall/2019 and Spring/2020 semesters.

## 3   Findings

Similar statements were combined during the refining of data. We found 7 main PROs and 9 main CONs in total; these statements are listed below.

The most important reasons considered as PROs for taking up a testing career among the surveyed individuals are presented in Table 2, along with their frequencies. The most frequent, with a 38% of respondents pointing that software testers earn more monetary benefits.

The second-most cited PROs, with 35%, is that testing tasks particularities allows software testers to learn about the system in hand, new technologies, and testing tools because the testing tasks provide a full background of the project scope, modularization, and integration strategy in a short period of time.

The third-most cited reason (31%) is the perception that software testing increases the software product quality thus worth doing despite the costs and efforts involved.

The fourth-most popular motivator is the perception that software tester deliver an important job, crucial to improve to guarantee that the software works properly and satisfies customer needs. Also, 12% of respondents mentioned job opportunities for software testers. Only a scarce 14% of individuals noted that software testing is a thinking job that requires critical thinking and creativity.

In contrast, when asked about the CONs for taking up a testing career, results are shown in Table 3. Respondents gave most importance to the following reasons: (a) it is a tedious job for 50% of respondents, (b) it is a complex task that requires expertise for 34%; and (c) 16% of the respondents pointed out the perception that other team members may become upset facing tester´s findings, i.e. software failures. Lastly, 7% of the subjects noted that in the labor market the role of tester is a role for which wages are lower than the average wages for other roles; that view is reinforced by the perception that software testers are treated as second-class citizens within a software project (6%).

Regarding the third question, Table 1 shows the responses to the actual chances of respondents taking up a testing career according to their personal preferences.

Table 1. Chances of students taking up software testing careers.

| Responses (132) | Numbers | Percentage |
|---|---|---|
| Certainly Not | 9 | 7% |
| No | 22 | 17% |
| May be | 70 | 53% |
| Yes | 24 | 18% |
| Certainly Yes | 7 | 5% |

## 3.1 The PROs of a Software Testing Career

The PROs responses are analyzed and presented in Table 2. Since we exclude PROs that were less than 5% (too small to consider) of the total, and the students could list up to three reasons, hence the total in the column is more than 100%.

Table 2. Percentage of salient PROs.

| PROs | Percentage |
|---|---|
| Learning opportunities | 35% |
| Important job | 26% |
| Easy job | 14% |
| Thinking job | 11% |
| More job opportunities | 12% |
| Monetary benefits | 38% |
| Increase product quality | 31% |

We analyzed the responses to our prompt and found some of the following principles to be common themes cited for PROs:

1. Learning opportunities – Testers can learn about different products, technologies, techniques, and languages as well as domains such as retail, finance, embedded systems, etc. Testing activities provide the full background of a project's scope and architecture in a short period of time and may span all project stages. Testers can also hone their softer skills due to more interactions with developers and customers, although many of them being difficult ones.
2. Important job – Testers are accountable and responsible for the product quality. In that sense, testing is perceived as an important part of the software life cycle.
3. Easy job – This refers to students' belief that testing does have well-defined and easy processes, etc. There is also a perception that testing does not require strong technical skills.
4. Thinking job - This encompasses views about testing such as being challenging, creative, innovative, and requiring logical and analytical thinking.

5. More job – This states that more testing jobs are available; due to the higher demands, the jobs are secure and stable.
6. Monetary benefits – Testing jobs pay good salary packages, as much or more than for software developers, thus it is a lucrative career.
7. Increase product quality – Some students felt that proper software testing will improve the quality of a software product, thus considering a worthwhile endeavor in terms of associated time and cost.

### 3.2 The CONs of a Software Testing Career

The CONs responses are analyzed and presented in Table 3. Since the students could list up to three CONs and we excluded CONs categories that were less than 5%, therefore the total percentages is more than 100%.

**Table 3.** Percentage of Salient CONs.

| CONs | Percentage |
|---|---|
| Second-class citizens | 6% |
| Career limitations | 16% |
| Complexity/Expertise needed | 34% |
| Tedious/Time consuming tasks | 50% |
| Prefer development | 12% |
| Less monetary benefits | 7% |
| Anti social/Find others' errors | 16% |
| Stressful job | 5% |
| Detailed-oriented skills | 5% |

We also analyzed the responses to our prompt and found some of the following principles to be common themes for CONs:

1. Second-class citizen – This points out that testers are not involved in decision-making. Testers often take the blame for poor quality software while developers reap the reward for good quality. They perceive an evident lack of support from management in terms of unrealistic schedules, poor allocation of resources, and inadequate recognition.
2. Career development – Students believe that there is limited growth in the testing field. Some also believe that testers' jobs are dead-end careers and that they are the first ones to lose their jobs during business downturns.
3. Complexity – Testers often face complex situations such as different versions of software products, platform incompatibilities, defects not getting reproduced, and testing not given sufficient time. At the same time, testers are often uniquely held responsible for product quality. This also includes the fact that testers need to look at business and technology artifacts and understand many abstractions. The lack of clarity around requirements adds to the difficulties. This factor also includes dealing with different versions and vendors of third party software, problems with testing tools, development environments or a weak infrastructure, and the ask exhaustively for patience.
4. Tedious – This refers to the repetitive nature of testing; respondents also used descriptors such as, monotonous, boring, not-challenging, and time-consuming.
5. Prefer development – Some students believe that they would miss opportunities for professional development by taking up testing careers. While some testers do develop test automation systems, the students seem to consider that as different from actual development activity. Some also think that they lose learning opportunities that are available to developers because they do not code anymore.
6. Less monetary benefits – Some students believe that testers' jobs do not have monetary benefits on par with developers.
7. Finding the mistakes of others – Fundamentally, it is not pleasant to find mistakes in others' work and present them. As testers have to report anything that adversely impacts the value of the product, which may not go well with some stakeholders; testers thus risk generating an anti-social role.
8. Stressful job – Students mentioned that software testers are gatekeepers of the software quality, therefore accountable for faulty software releases. Some think that tester must be on stand-by all the time, just in case a software fails and actions must be taken, thus negatively affecting the work-life balance.
9. Detailed-oriented skills – Some students just mentioned that they have no interest in the field of software testing because it requires concentration for several hours, every day, looking for details or edge conditions to test software.

## 4 Threats to Validity

This study has some inherent limitations. The subjects come from two large public universities and one top-ranked private university in the United Arab Emirates only. Although respondents represent a sample of currently active senior students throughout the UAE, including expatriates, their origins were not recorded.

## 5 Discussions and Implications

The study analyzed the opinions of 132 senior students enrolled in information science, information and communication technology, computer science and software-related disciplines in the UAE. Opinions were elicited on whether they would choose testing careers and what they felt were the advantages and drawbacks (PROs and CONs) of a testing career.

We found considerable variations in the student-perceived advantages of a career in testing. Among the key PROs of learning opportunities, monetary benefits, and job opportunities, percentages narrowly vary from 31% to 38%. With 18% of students indicating that they would like to be a software tester, and only 6% saying they would certainly take up a career as software tester, it is evident that a software testing profession among students is as unpopular as it is in other countries [19]; where only 7% of the students would not like to be software tester. These results do concur with prior studies [20] [21], which point to the tester role as one of the least popular roles compared to others, such as project manager, analyst, designer, programmer, and maintenance.

Not surprisingly, there are few outliers found in this investigation that are worth mentioning: (a) five respondents saw in software testing an opportunity to beat or challenge the developer; (b) stressful job was mentioned by only 5% of the subjects in this study, consistent with previous studies where this issue appeared less than 5%; and (c) five students alarmingly asserted that software testing is too expensive, and a waste of time and effort because the tester may not find any error in the software. Such students have an alarming misconception about the role of testing in the software life cycle, and surely need to take a course on software verification and validation.

This study has many implications for colleges, especially for information technology, information science, computer science, software engineering programs, and information and communication technology programs. Since testing courses can improve the perception of testing careers, universities can introduce them into their curricula. They can regularly review the curricula by consulting their alumni and researchers. Since testing offers additional jobs, a testing course can help colleges improve placement prospects of their students.

The program curricula, software testing and verification and validation courses need to reflect the understanding that testers need to provide correct information to various stakeholders, and appreciate that testing is 'applied epistemology' grounded in 'social psychology'. Course instructors must dispel beliefs such as, 'testing is just mechanically running tests and comparing outputs with expected results'. Instead, testing instructors should explain the importance of testing and the philosophy behind it, and they should impress upon the students that any testing assignment and design of test cases can be very creative.

Instructors should educate tester who can understand different domains and the needs of users in these different domains. These testers must be able to understand the developer mindset, and to anticipate mistakes that developers may make. These testers must do their testing jobs creatively and efficiently under challenging constraints, and they must report finding wisely and tactfully to all stakeholders.

## 6 Conclusions

As previously emphasized, software testing is a human activity [22] [23] and testers, who willingly take up testing careers, can influence the quality of the final product.

Nevertheless, software testing appears to be a neglected area in the software industry [24]. There are not enough testing specialists, and test schedules are squeezed as development overruns occur and delivery milestones are considered non-negotiable. Many times, testing is perceived as a nuisance that is sandwiched between development and deployment, when in fact it is a critical activity that needs to be performed in parallel to design and development activities, as advocated by software verification and validation. Therefore, the software industry has been facing a shortage of qualified software testers which has led to quality problems.

This study offers useful insights and helps educators to come up with an action plan to change the outlook towards testers in their programs; educators must put the software testing profession under a new light. Enthusiasm for software testing could increase the number of software engineers deciding on testing as a career of their choice, could increase the quality of software testing, and improve the overall productivity and turnaround time of software development activity.

## Appendix A – Survey Questions

1. What are the three PROs (in the order of importance) for taking up testing career?
    a)                b)                c)
2. What are three CONs (in the order of importance) for taking up testing career?
    a)                b)                c)
3. What are chances of my taking up testing career?
    Certainly Not   No   Maybe   Yes   Certainly Yes
    Reasons:
4. Gender (optional):
5. GPA (optional):